**Meta-analysis of movements in Atlantic leatherback turtles during nesting season: conservation implications**

Jean-Yves Georges[1*], Alexis Billes[2], Sandra Ferraroli[1,3], Sabrina Fossette[1,3], Jacques Fretey[4], David Grémillet[1,5], Yvon Le Maho[1], Andrew E. Myers[6], Hideji Tanaka[7], Graeme C. Hays[6]

[1] Centre National de la Recherche Scientifique, Institut Pluridisciplinaire Hubert Curien, Département Ecologie, Physiologie et Ethologie, Unité Mixte de Recherche 7178 CNRS-Université Louis Pasteur, 23 rue Becquerel, 67087 Strasbourg, France

[2] Programme Kudu, Cellule de coordination ECOFAC, BP 15115, Batterie IV, Libreville, Gabon

[3] Université Louis Pasteur, 4 rue Blaise Pascal, 67070 Strasbourg, France

[4] Union Internationale pour la Conservation de la Nature, Museum National d'Histoire Naturelle, 36 rue Geoffroy Saint-Hilaire, 75005 Paris, France

[5] Percy FitzPatrick Institute, DST/NRF Centre of Excellence, University of Cape Town, Rondebosch 7701, South Africa

[6] School of Biological Sciences, Institute of Environmental Sustainability, University of Wales Swansea, Singleton Park, Swansea SA2 8PP, UK

[7] COE for Neo-Science of Natural History, Graduate School of Fisheries Sciences, Hokkaido University, 041-8611 Hakodate, Japan

* Corresponding author:  Jean-Yves Georges

Phone: +33 388 106 947

Fax: +33 388 106 906

Email: jean-yves.georges@c-strasbourg.fr

**Running title:** inter-nesting movements in Atlantic leatherbacks

**Abstract**

Despite decades of conservation efforts on the nesting beaches, the critical status of leatherback turtles shows that their survival predominantly depends on our ability to reduce at-sea mortality. Although areas where leatherbacks meet fisheries have been identified during the long distance movements between two consecutive nesting seasons, hotspots of lethal interactions are still poorly defined within the nesting season, when individuals concentrate close to land. Here we report movements of satellite-tracked gravid leatherback turtles during the nesting season in Western Central Africa, South America and Caribbean Sea, accounting for about 70% of the world population. We show that during, and at the end of, the nesting season leatherback turtles have the propensity to remain over the continental shelf, yet sometimes perform extended movements and may even nest in neighbouring countries. Leatherbacks exploit coastal commercial fishing grounds and face substantial accidental capture by regional coastal fisheries (e.g. at least 10% in French Guiana). This emphasises the need for regional conservation strategies to be developed at the ocean scale, both at sea and on land, to ensure the survival of the last leatherback turtles.

**Introduction**

Understanding how marine animals interact with human activities is urgently needed to devise sound management strategies for marine ecosystems threatened by climate change (e.g. Parmesan & Yohe 2003) and direct anthropogenic pressure (Lewison et al. 2004). Satellite telemetry has provided evidence that long-distance migrating species of commercial interest can perform pan-oceanic movements regardless of country delimitations (e.g. Lutcavage et al. 1999). It has also been showed that fishing fleets aggregate in the same areas, thereby increasing the probability of incidental catch of species of conservation interest, such as endangered sea turtles, both in oceanic and coastal waters (e.g. Epperly 2003, Ferraroli et al. 2004, James et al. 2005).

While satellite telemetry is being widely used to document the large scale displacements of leatherback sea turtles *Dermochelys coriacea* between two consecutive nesting seasons (e.g. Morreale et al. 1996, Ferraroli et al. 2004, James et al. 2005, Eckert 2006), our knowledge of their movements during the nesting season remains extremely limited. Initially, recovery of flipper tags used on nesting leatherbacks for demographic studies incidentally showed that gravid females can move among regional nesting beaches (e.g. Eckert et al. 1989a, Girondot & Fretey 1996). Later studies involving radio-telemetry showed that leatherback females from Malaysia, South China and Caribbean disperse 30 to 40 km from the coastline on the continental shelf, covering up to 300 km between two consecutive nesting events (Chan et al. 1991, Eckert et al. 1989b, 1996). Such telemetry studies sometimes contributed to the establishment of time-area closures and gear modifications that successfully helped the conservation of this species (e.g. Chan et al. 1991, Hays 2004). Yet the precise movements of leatherbacks between two consecutive nesting events remain largely unknown in most nesting areas worldwide (but see Eckert et al. in press, Fossette et al. in press).

Interactions of leatherback turtles with coastal fisheries are potentially frequent when females and males congregate close to nesting sites around the nesting season (Chan et al. 1991, James et al. 2005, Fossette et al. in press) but also in some cases after the nesting season (Eckert et al. in press). Indeed, accidental catches of sea turtles in coastal and shelf waters have recently been identified as a rapidly emerging threat, particularly for the critically endangered leatherback turtle (FAO 2004). It is therefore crucial not only to know where leatherbacks congregate between two consecutive nesting events, but also to detail their at-sea movements during the nesting season. This is necessary to understand better how leatherbacks interact with fisheries and ultimately to assess the magnitude of these interactions.

We examined the movements of satellite-tracked female leatherback turtles during (inter-nesting tracks) and immediately after (post-nesting tracks) the nesting season in Western Central Africa, South America, and the Caribbean Sea, respectively accounting for about 70% of the world population of leatherbacks since their precipitous decline in the Pacific Ocean (Troëng et al. 2004). The three study sites have coastal characteristics ranging from the wide and narrow continental shelf in South America and Western Africa, respectively, to islands surrounded by deep waters in the Caribbean Sea. Since leatherbacks from South China and the Caribbean have been reported to adjust their at-sea behaviour according to local bathymetry (Chan et al. 1991, Eckert et al. 1989b, 1996), we here aim to identify general patterns and potential differences in leatherback movements within the nesting season according to coastal specificity at the Atlantic scale. In the Western Atlantic, the immediate proximity of substantial populations nesting in neighbouring countries, and the propensity of this species for straying (i.e. in travelling between nesting areas during an inter-nesting interval, see Eckert et al. 1989a) suggest that leatherback turtles from this zone may have widespread movements that take them to other countries between two consecutive nesting events. Here we used satellite telemetry to investigate the movements of leatherback turtles between two consecutive nesting events in order to understand better how potential transbordering movements occur, and to identify potential hot-spot areas where they are more likely to interact with fisheries operating in the area.

**Methods**

Inter-nesting movements of leatherback turtles were investigated using satellite transmitters (French Guiana and Grenada: Series 9000X Satellite Relay Data Loggers, by Sea Mammal Research Unit, UK: 350 g in air, 10.5 cm long, 28 cm² section; Gabon in 2002: KiwiSat 101 satellite transmitters, by Sirtrack Ltd, New Zealand: 650 g in air, 18 cm long, 20 cm² section, and Gabon in 2003: SDR-T16 satellite transmitters, by Wildlife Computer, USA: 380 g in air, 10 cm long, 35 cm² section) and reconstructed using the Argos system (http://www.argosinc.com/). Transmitters were deployed on 9 females nesting in 2002 (n=5) and 2003 (n=4) at Mayumba beach (3.7°S, 10.9°E) on the south coast of Gabon, 13 females nesting in 2002 (n=3) and 2003 (n=10) at Levera beach (12.2°N, 61.6°W) on the north shore of Grenada, and 10 females nesting in 2004 at Awala Yalimapo beach (5.7°N, 53.9°W) on the western border of French Guiana with Suriname. In all sites, transmitters were deployed when female turtles had completed their nesting event and were held on the carapace using a customised flexible harness (Fossette et al. submitted).

Transmitters were equipped with a salt-water switch that relayed the schedule of time spent in or out of the water. Accordingly, we defined a haul-out as any period >10 minutes when the switch was continuously dry. Based on the tracking and haul-out data, along with direct observation in some cases, we inferred whether turtles re-nested after transmitter attachment in order to categorise tracks as either inter-nesting or post-nesting tracks, but also to identify potential by-catch of satellite-tracked animals. Movement data were analysed using all Argos-derived positions classified as 1, 2 or 3 (with nominal standard deviations around the true position of 150m, 350m and 1000m, respectively), excluding locations implying a velocity above 10 km/h (Fossette et al. in press). We then calculated the mean daily position for each turtle as the means of equally weighed latitudes and longitudes obtained for a given day, to provide a simplified representation of movements and to avoid pseudo-replication.

**Results**

All 9 turtles from Gabon in 2002 and 2003, all 10 turtles from French Guiana in 2004 and 5 of the 10 turtles tagged from Grenada in 2003 were monitored at sea during at least one inter-nesting trip, spending 10.7 ± 1.0 days (mean ± SD, n=9 complete inter-nesting trips), 10.3 ± 2.2 days (n=8 complete trips), and 13.8 ± 3.2 (n=5 complete trips) days, at sea between two consecutive nesting events in Gabon, French Guiana, and Grenada, respectively (**Table 1**). In all 3 locations, leatherbacks moved away from the departure point by 102 ± 50 km off Gabon, 121 ± 51 km off French Guiana, and 134 ± 37 km off Grenada and yet travelled hundreds of km between two consecutive nesting events (656 ± 144 km from Gabon, 560 ± 134 km from French Guiana, and 384 ± 116 km from Grenada; **Table 1**). Importantly, leatherback turtles visited waters of neighbouring countries, namely the Republic of Congo for Gabon (**Figure 1a**), Surinam for French Guiana (**Figure 1b**), and St Vincent Island for Grenada (**Figure 1c**).
In Gabon, all but one (GA06) turtle remained on the shallow continental shelf (<200 m deep, **Figure 1a**), three of them (GA04-05-08) remaining in front of the nesting beach within 60 km of the coast. All other individuals moved parallel to the coast, two of them (GA01-09) heading north in Gabon waters and moving clockwise before returning to the beach.

Conversely, the four other turtles (GA02-03-06-07) headed South to Congo waters where they moved anticlockwise, with turtle GA06 reaching deep waters (>200m) off the Congo. Despite this wide dispersion, turtles spent in total 47% of their time in shallow waters within 40 km of Mayumba beach, and landed 5.0 ± 2.3 km (range 1.7 – 9.8 km) away from their departure point (**Table 1**).

Leatherbacks in French Guiana behaved in a similar manner to those in Gabon. All turtles remained on the shallow continental shelf (<200 m deep, **Figure 1b**), with six individuals reaching waters deeper than 50 m (FG01-02-03-04-05a-07a), and the others remaining within 60 km of Awala-Yalimapo beach, where waters are approximately 20 m deep. Six individuals (FG03-04-05-06-07-08) remained in French Guiana waters East of Awala-Yalimapo beach and moved clockwise before returning to land. Conversely, the four other turtles headed West into Surinam waters where they moved anticlockwise. Such a pattern, also observed in Gabon, permits returning turtles to remain close from the shore. Turtles spent in total 40% of their time at sea within a 20-km radius area in shallow waters (<50 m deep) in front of the Maroni River estuary, on the border between French Guiana and Surinam and landed 7.2 ± 6.0 km (range 1.8 – 19.8 km) away from their departure point (**Table 1**).

In Grenada, all five leatherbacks monitored for their complete inter-nesting trips remained West of the island in deep Caribbean waters and did not occur on the shallow peri-insular shelf on the Atlantic side, East of Grenada (**Figure 1c**). All turtles initially headed North from Levera beach before moving either clockwise (n=3 inter-nesting trips) or anticlockwise (n=3 inter-nesting trips) West of a NE-SW line joining the South coast of St Vincent to the North coast of Grenada (where Levera beach lies). All five turtles returned to Grenada on Levera beach, or beaches nearby for their subsequent nesting event (**Table 1**).

The three turtles monitored in 2002 and seven of the ten turtles studied in 2003 in Grenada were also tracked during the initial phase of their post-nesting movements (**Figure 2**). They all remained for several days within the Caribbean Sea before entering the Atlantic Ocean. The most important point is that individual turtles used different routes when entering the Atlantic, passing through almost all channels of the Western Indies Arc. For example, GR12 and GR13 entered the Atlantic by passing between Grenada and St Vincent, i.e. 100 km from Levera beach, whereas GR11 travelled 600 km within the Caribbean Sea before entering the Atlantic Ocean in the vicinity of St Croix.

**Discussion**

The present tracking of leatherback turtle movements during the nesting season on both sides of the Atlantic Ocean has identified consistent extended movements between two consecutive nesting events and a wide use of coastal areas ranging from shallow continental shelf and slope waters to deep peri-insular waters throughout this ocean.

The different types of marine habitats used by leatherbacks in Grenada compared to French Guiana and Gabon result from contrasting bathymetric and hydrologic conditions between sites. Grenada is a volcanic island characterised by a narrow peri-insular shelf dropping down to 1000m within 20 km on the Atlantic side, and down to 2000 m within 10 km in the Caribbean side (**Figure 1c**). Atlantic surface waters generally flow into the Caribbean between all islands in the Lesser Antilles and exit via the Yucatan Channel, with surface currents being massively faster through the most southern passages (Windward Islands Passages Monitoring Program http://www.aoml.noaa.gov/phod/wimp/caribb1.html). Such strong currents may prompt leatherbacks from Grenada to remain within the deep Caribbean waters rather than entering the Atlantic. Conversely, Gabon and French Guiana have a continental shelf 50-km and 100-km wide, respectively (**Figures 1a, 1b**) both influenced by slower surface currents (Froidefond et al. 2002). Accordingly, leatherbacks can disperse broadly above the continental shelf of Gabon and French Guiana, incurring basically no constraints from these weak currents, their dispersion range at sea being ultimately restricted by the timing of egg-maturation. Our results confirm previous studies showing that leatherbacks adjust their at-sea behaviour according to local bathymetry. For instance, previous studies reported leatherbacks performing deep dives in deep Caribbean waters around St Croix, US Virgin Islands (Eckert et al. 1989b) contrasting with thir shallow diving behaviour on shallow continental shelf in South China (Eckert et al. 1996). This indicates that leatherback turtles are primarily deep divers unless physical constraints such as bathymetry prompt them to restrict their diving performances.

In our study, leatherbacks from Gabon and French Guiana travelled distances that were, on average, 30-40% longer than in Grenada in 15-20% less time, while the overall mean dispersion range differed by only 5-20 km. Such differences may result from leatherbacks diving deeper (Hays et al. 2004), and hence moving shorter horizontal distances in Grenada compared to Gabon and French Guiana where they mostly move horizontally by performing shallow dives (Fossette et al. in press, for French Guiana). The inter-nesting interval durations ranged from 9 to 18 days, as previously reported from tag recaptures (e.g. Girondot & Fretey 1996). Considering that leatherback turtles lay on average 6 to 7 consecutive nests per season (e.g. Girondot & Fretey 1996), the inter-nesting movements reported in our study indicate that a leatherback can move from 2000 to 4500 km during the entire nesting season depending on the nesting site. Clearly leatherbacks do not simply rest close to their nesting beach and the observed extended movements might be associated with mating or foraging. Recent video data obtained in Costa Rica show that leatherback females avoid interactions with males by lying motionless or moving slowly (Reina et al. 2004). The extended movements presented in our study may not be interpreted as sexual interactions. With regards to foraging, direct and indirect evidence supports the hypothesis that extended movements may be associated with a search for food. Leatherback turtles feed primarily on gelatinous plankton that often forms part of a deep scattering layer in the oceans moving towards the surface at dusk and descending again at dawn (Davenport 1998). In both French Guiana and the Caribbean Sea, leatherbacks have been reported to dive extensively during the nesting season, showing a diurnal pattern (Eckert et al. 1989b, Hays et al. 2004, Fossette et al. in press) consistent with feeding within deep scattering layers. Complementary behavioural data, including diving and feeding events, are required, particularly for Gabon, to confirm the hypothesis of leatherback turtles feeding during the nesting season.

Our results also show interesting consistencies among the study sites and consequently raise common key conservation issues both at sea and on land for the Atlantic populations of leatherback turtles. During their extensive inter-nesting movements at sea, leatherbacks tend to remain on the continental shelf, when present, but utilise Economic Exclusive Zones (EEZs) from different countries. Most countries maintain these mandatory EEZs up to 200km from their shore, within which they control fisheries practices. Four of the nine leatherback turtles tracked from Gabon visited Congo waters and moved close from the shore when returning to their nesting site (**Figure 1a**). Importantly, illegal trawl-net fisheries are regularly observed from the shore operating between the coast of Mayumba and the border with Congo (AB, JF & JYG personal observation). Such uncontrolled fisheries may have dramatic impact on sea turtles, including leatherbacks but also green *Chelonia mydas* and olive Ridley *Lepidochelys olivacea* turtles, although this has never been quantified in Western Africa. In French Guiana, leatherbacks spent ~40% of their time at sea on the border between French Guiana and Surinam, with 40% of them moving into Surinam waters. The coastal area is heavily exploited by 2 main local fisheries which target either brown shrimps *Penaeus subtilis* or red snappers *Lutjanus purpureus* (Ferraroli et al. 2003, **Figure 3**). These fishing practices occur throughout the nesting season (Fabian Blanchard, pers. comm.) and put leatherback turtles under substantial threat (Ferraroli et al. 2003) as can be seen when assessment of the data from turtle FG07: After 3 consecutively-recorded inter-nesting trips, this turtle was apparently pulled out of water within French Guiana EEZ (6°N – 54°W). Its transmitter then remained within French Guiana waters for a few days before it moved straight to Georgetown, Guiana, at a constant high speed of 13 km/hour over 24 hours, indicating that FG07 has been captured by a vessel hauling out in Guiana (**Figure 1b insert**). Such data highlight the threat faced by leatherbacks due to foreign fisheries operating, sometimes illegally (Ferraroli et al. 2003), in French Guiana waters. One may expect that such by-catch may also result from legal local fisheries (Ferraroli et al. 2003). In Grenada, the transmitter attached to turtle GR02 came out off the water and was then located for several weeks in a village on St Vincent. We did not get any return information regarding the circumstances of these turtles' captures, nor about their possible release. However, both examples clearly show that satellite tracking can add valuable information to turtle accidental catch and mortality which complement onboard direct observation. Additionally, despite the substantial site fidelity shown here, some individuals from French Guiana routinely visit different beaches and sometimes neighbouring Suriname probably for nesting.

For some sea turtle populations remaining close to their nesting beach during the breeding season and feeding in narrow foraging grounds, focused conservation efforts on restricted areas can prove tremendously successful (e.g. Hays 2004). Our investigation shows that this is clearly not the case for leatherback turtles, which can move far away from nesting beaches during a single inter-nesting interval, frequently crossing international borders and also nesting on the shore of neighbouring countries. A recent study reported leatherbacks nesting in Florida to remain on the Atlantic side of the continental shelf even after the nesting season, yet frequenting waters of different states of the USA (Eckert et al. in press). In the Caribbean Sea, leatherbacks initiate their post-nesting movements by using indirect routes that take them close to a wide range of countries. Altogether, this study demonstrates that leatherback turtles need to be considered as shared resources at sea but also on land. This may be achieved by expanding conservation initiatives to act on a regional scale, as opposed to a national scale during the nesting season (see e.g. Chan et al. 1991), and beyond, in order to assess population trends accurately, define trans-bordering protected areas both marine and terrestrial, and concertedly control international fishing practices if the survival of the last large population of leatherback turtles is to be secured.

**Acknowledgments**

We are grateful to all concerned public services, particularly the Gabonese Forestry Commission and the National Council of the Gabonese National Parks, the Ministry of Ecology and Sustainable Development and the Direction Régionale de l'Environnement-Guyane in French Guiana, and the Ministry of Agriculture, Forestry, Land and Fisheries in Grenada. We thank all participants of sea turtle monitoring programmes developed in Mayumba beach (Nyamu NGO and Gabon-Environnement NGO), Awala-Yalimapo beach (Réserve Naturelle de l'Amana, Kulalasi, and Tom Doyle) and Levera beach (Ocean Spirits Inc. and Jon Houghton) for logistical help in the field. Thanks to F Blanchard from Institut Français pour la Recherche pour l'Exploitation de la Mer, IFREMER-Guyane) for the information related to fisheries activities in French Guiana. Funding was provided by grants to A Billes and J Fretey from European ECOFAC Program, J-Y Georges and Y Le Maho from the European FEDER Program, and to GC Hays from the Natural Environment Research Council of the UK (NERC). S Fossette and AE Myers were supported by a studentship from the French Ministry of Research and NERC, respectively. This study was performed in accordance with institutional, national and international guidelines and authorisations for the use of endangered species in research.

**Figure 1** Summary of the inter-nesting movements performed by Argos tracked leatherback turtles nesting in (a) Gabon, (b) French Guiana, and (c) Grenada, in relation to bathymetry. Arrows indicate direction of travel with turtle number shown adjacent (a, b and c denote successive inter-nesting tracks by the same individual). (a) Insert: location of the three study sites. (b) Insert: track of the transmitter fitted to FG07 being caught during its fourth inter-nesting trip (FG07d).

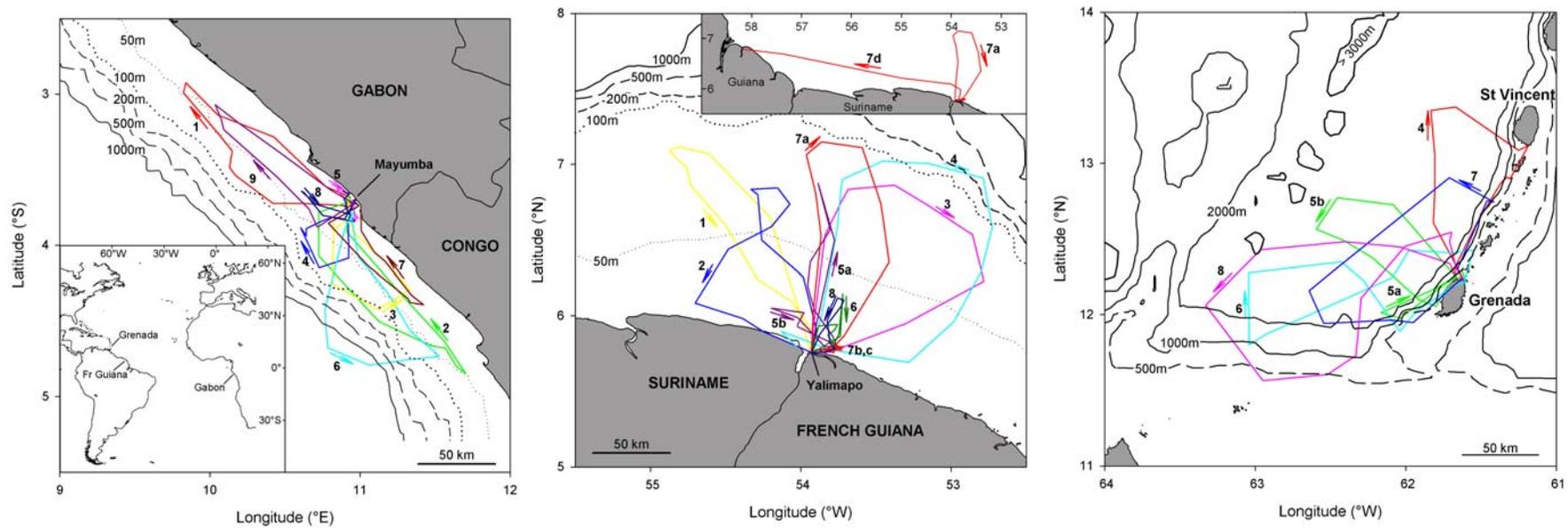

**Figure 2** Post-nesting movements of 10 Argos tracked leatherback turtles nesting on Grenada in 2002-2003 in relation to Exclusive Economic Zones in the Caribbean region (in dashed lines). The track of turtle GR02 (in green) stopped at St Vincent.

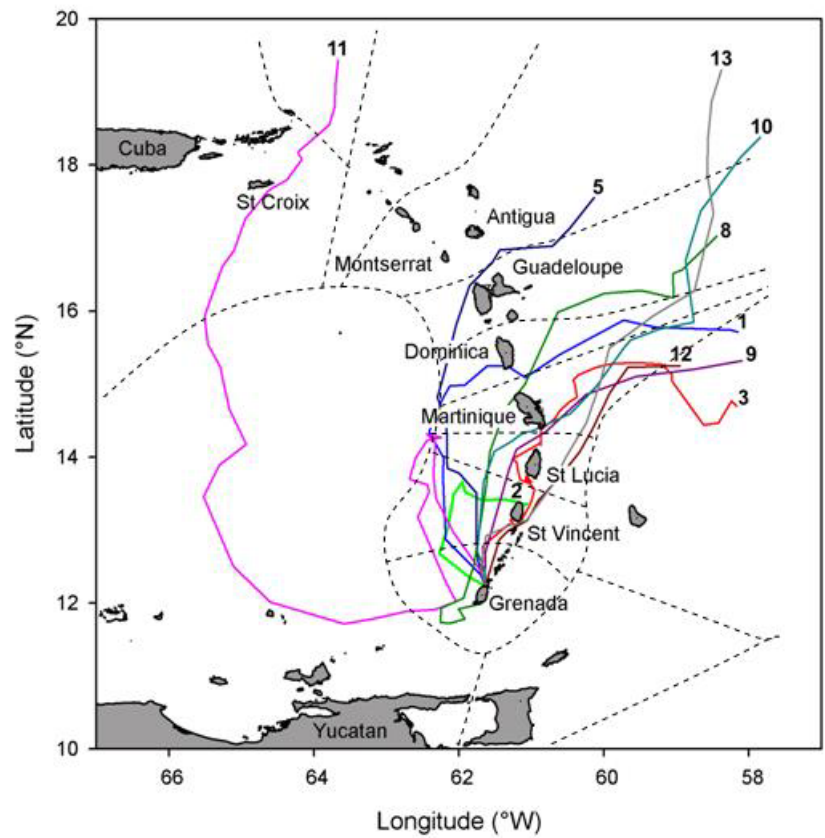

**Figure 3** Topographic representation of space use by 10 Argos-tracked leatherback turtles during their inter-nesting movements off French Guiana in relation to bathymetry and local fisheries targeting red snapper *Lutjanus purpureus* (red) and brown shrimp *Penaeus subtilis* (green). The blue colour denotes the total time (in days) turtles spent in each 0.1° x 0.1° square. Note that the fisheries distribution only relates to French Guiana (from Ferraroli et al. 2003).

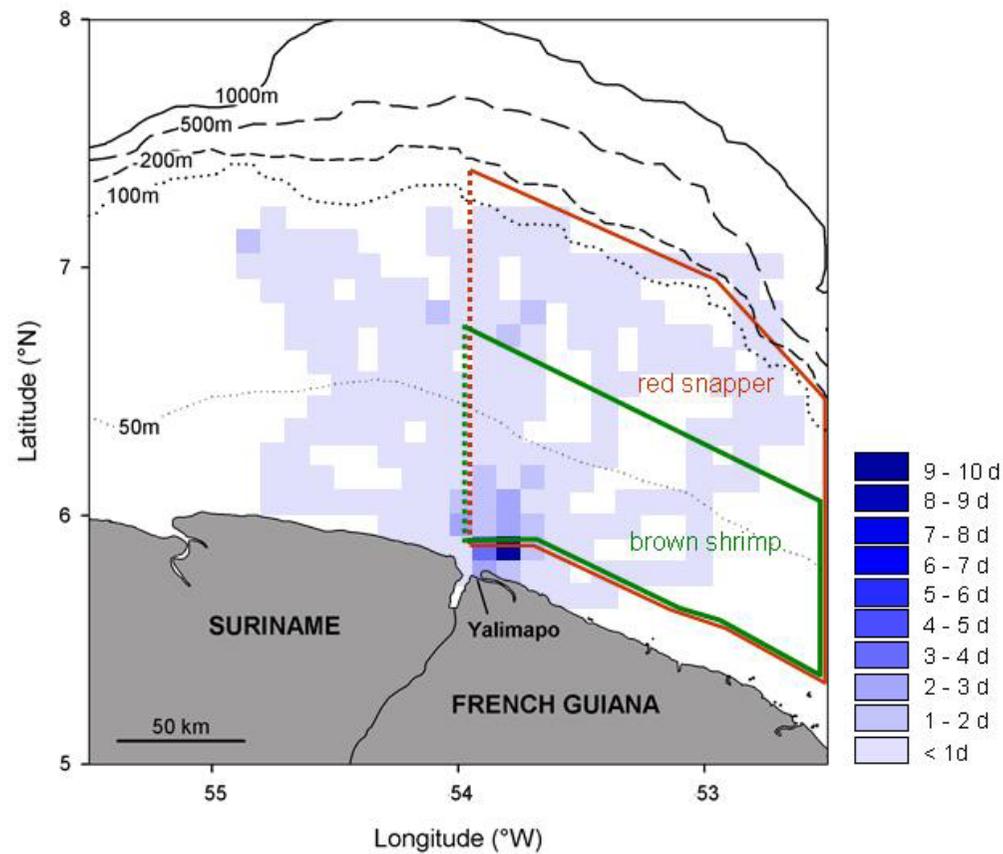

Table 1 Individual summary of the inter-nesting movements performed by Argos tracked leatherback turtles nesting in Gabon (GA#), French Guiana (FG#), and Grenada (GR#). a, b and c denote successive trips by the same individual. Mean ± SD values were calculated for each site, and for all sites, considering mean individual values for turtles being monitored over successive trips. ANOVAs were performed to test means among locations.

| Turtle ID | Date of departure | Trip duration (days) | Total distance travelled (km) | Max distance from tagging location (km) | Distance to second landing (km) |
| --- | --- | --- | --- | --- | --- |
| GA01 | 19 Dec 2002 | 12.2 | 879 | 165 | 3.0 |
| GA02 | 20 Dec 2002 | 11.1 | 760 | 163 | 1.9 |
| GA03 | 17 Dec 2002 | 10.9 | 446 | 88 | 7.9 |
| GA04 | 24 Dec 2002 | 11.9 | 477 | 59 | 3.5 |
| GA05 | 10 Dec 2002 | 10.0 | 687 | 34 | 5.3 |
| GA06 | 11 Dec 2003 | 10.2 | 742 | 136 | 3.5 |
| GA07 | 11 Dec 2003 | 9.3 | 568 | 97 | 7.2 |
| GA08 | 12 Dec 2003 | 9.4 | 592 | 45 | 9.8 |
| GA09 | 11 Dec 2003 | 11.2 | 756 | 133 | 3.7 |
| FG01 | 15 May 2004 | 10.4 | 671 | 189 | 19.8 |
| FG02 | 17 May 2004 | 12.2 | 677 | 136 | 2.7 |
| FG03 | 16 May 2004 | 10.0 | 578 | 168 | 2.8 |
| FG04 | 16 May 2004 | 12.0 | 750 | 191 | 12.3 |
| FG05a | 17 May 2004 | 15.0 | 809 | 135 | 2.1 |
| FG05b | 1 June 2004 | 12.1 | 780 | 53 | 5.7 |
| FG06 | 14 May 2004 | 11.4 | 423 | 56 | 1.8 |
| FG07a | 15 May 2004 | 11.1 | 699 | 163 | 19.4 |
| FG07b | 26 May 2004 | 12.0 | 388 | 35 | 6.7 |
| FG07c | 7 June 2004 | 9.9 | 378 | 34 | 6.2 |
| FG07d | 17 June 2004 | † | † | † | † |
| FG08 | 16 May 2004 | 8.7 | 370 | 61 | 3.3 |
| FG09 | 17 May 2004 | $5.3^1$ | $394^1$ | $141^1$ | - |
| FG10 | 15 May 2004 | $9.4^1$ | $579^1$ | $131^1$ | - |
| GR01 | 4 July 2002 | $-^2$ | $-^2$ | $-^2$ | $-^2$ |
| GR02 | 7 July 2002 | † | † | † | † |
| GR03 | 9 July 2002 | -² | -² | -² | -² |
| GR04 | 30 March 2003 | 18.0 | 343 | 127 | Levera beach |
| GR05a | 30 April 2003 | 10.0 | 153 | 65 | Levera beach |
| GR05b | 11 May 2003 | 10.0 | 282 | 114 | Levera beach |
| GR06 | 5 April 2003 | 16.0 | 460 | 163 | Levera beach |
| GR07 | 26 April 2003 | 12.0 | 378 | 112 | Levera beach |
| GR08 | 18 May 2003 | 13.0 | 520 | 180 | Levera beach |
| GR09 | 11 May 2003 | -² | -² | -² | -² |
| GR10 | 8 July 2003 | -² | -² | -² | -² |
| GR11 | 14 June 2003 | -² | -² | -² | -² |
| GR12 | 17 June 2003 | -² | -² | -² | -² |
| GR13 | 27 June 2003 | -² | -² | -² | -² |
| | | | | | |
| **GAall** | - | **10.7 ± 1.0** | **656 ± 144** | **102 ± 50** | **5.1 ± 2.6** |
| **FGall** | - | **10.3 ± 2.2** | **560 ± 134** | **123 ± 51** | **7.1 ± 6.5** |
| **GRall** | - | **13.8 ± 3.2** | **384 ± 116** | **134 ± 37** | - |

| | | | | |
|---|---|---|---|---|
| **ANOVAs (F, P values)** | $F_{2,21} = 4.959, P = 0.017$ | $F_{2,21} = 6.549, P = 0.006$ | $F_{2,19} = 0.818, P = 0.455$ | $F_{1,15} = 0.715*, P = 0.411*$ |
| **Pooled data** | 11.6 ± 1.6 | 533 ± 113 | 120 ± 13 | 6.1 ± 1.0* |

[1] incomplete records

[2] post-nesting movements

† incomplete records due to accidental catch

* for GA and FG only